\documentclass[preprint2,longabstract]{aastex}

\newcommand{\muas}{$\mu{\rm as}$}
\slugcomment{Submitted to PASP}
\shorttitle{Crowded-Field Astrometry with SIM - II}
\shortauthors{R.\ Sridharan \& R.J.\ Allen}

\begin{document}


\title{Crowded-Field Astrometry with \textit{SIM PlanetQuest}. \\[0.1in]
II. An Improved Instrument Model \\[0.2in]
}

\author{\sc{R.\ Sridharan and Ronald J. Allen}}
\affil{Space Telescope Science Institute, 3700 San Martin Drive, 
Baltimore, MD 21218}
\email{sridhar@stsci.edu, rjallen@stsci.edu}

\begin{abstract}
In a previous paper we described a method of estimating the single-measurement
bias to be expected in astrometric observations of targets in crowded fields
with the future Space Interferometry Mission (SIM). That study was based on a
simplified model of the instrument and the measurement process involving a
single-pixel focal plane detector, an idealized spectrometer, and continuous
sampling of the fringes during the delay scanning. In this paper we elaborate on
this ``instrument model'' to include the following additional complications:
spectral dispersion of the light with a thin prism, which turns the instrument
camera into an objective prism spectrograph; a multiple-pixel detector in the
camera focal plane; and, binning  of the fringe signal during scanning of the
delay. The results obtained with this improved model differ in small but
systematic ways from those obtained with the earlier simplified model. We
conclude that it is the pixellation of the dispersed fringes on the focal plane
detector which is responsible for the differences. The improved instrument model
described here suggests additional ways of reducing certain kinds of confusion,
and provides a better basis for the evaluation of instrumental effects in the
future. %
\end{abstract}


\section{Introduction}

\textit{The Space Interferometry Mission PlanetQuest} (hereafter SIM) is being
designed by NASA/JPL to carry out a program of extremely precise astrometry on
stars, thereby contributing to a wide variety of research topics in astronomy
including the study of the mass distribution in the Galaxy and the search for
earth-like extra-solar planets. A few of these topics involve making precise
position measurements of target stars in areas of the sky containing more than
$\approx\, 0.4$ detectable stars per square arcsecond. In these
``crowded-field'' cases we have previously shown \cite[hereafter Paper
I]{sriron07a} that the individual astrometric measurements made on such targets
could be biased by the light from extraneous field stars. In Paper I of this
series, we introduced a simplified instrument model for SIM, and used it to
evaluate the likelihood of confusion bias on a number of Key Projects to be
carried out as part of the SIM science program. In that paper we concluded that,
in the small number of cases where confusion bias may be problematic, the
likelihood of bias can be reduced by judicious design of the observing program.

The measurement model used in Paper I is based on a phasor description for
estimating the contributions of all stars in the field of view (FOV). This part
of our model is not in question here. However, there were three subsequent
simplifications to the instrument model which were suspect: First, the detector
in the focal plane was assumed to consist of one large pixel per channel which
collected all the light diffracted through the field-stop; second, the details
of just how the light is dispersed in wavelength were ignored; and third, the
total fringe signal was assumed to be continuously sampled during the scanning
of the delay. In this paper we elaborate further on the instrument model to
include the following complications: A multiple-pixel detector in the focal
plane; spectral dispersion of the light with a thin prism, which effectively
turns the instrument into an objective prism spectrograph; and, integration and
discrete sampling of the fringe signal during scanning of the delay. We then
repeat the computations of confusion bias for a few specific cases already
evaluated with the simplified approach of Paper I. In general, the improvements
to the instrument model introduced in this paper have only a minor impact on the
biases, and come at the cost of significant complication. However, we
have not yet implemented a number of instrumental effects which may be present
in real SIM data, including pointing errors, delay line jitter, channel
band-shape changes, and so on. Besides its pedagogical value, the more complete
description of how SIM works which we give here will be useful for such studies
in the future.

We begin with stepwise description of the operation of SIM, progressively adding
complexity at each step. This leads us to the concept of the \textit{fan
diagram}, which was introduced in the context of SIM in Paper I as an aid to
understanding how SIM works, and which will gain further in importance here. We
then use this improved instrument model to re-evaluate the confusion bias for
several cases presented in Paper I. Finally, we point out a new possibility
related to the pixellation of the focal plane which may be used to further
reduce the likelihood of confusion bias from field stars located within the SIM
FOV but separated by $\gtrsim 1''$ from the target star.

\section{SIM Beam combination, dispersion, and focal plane imaging}

\begin{figure*}[!ht]
\epsscale{1.5}
\plotone{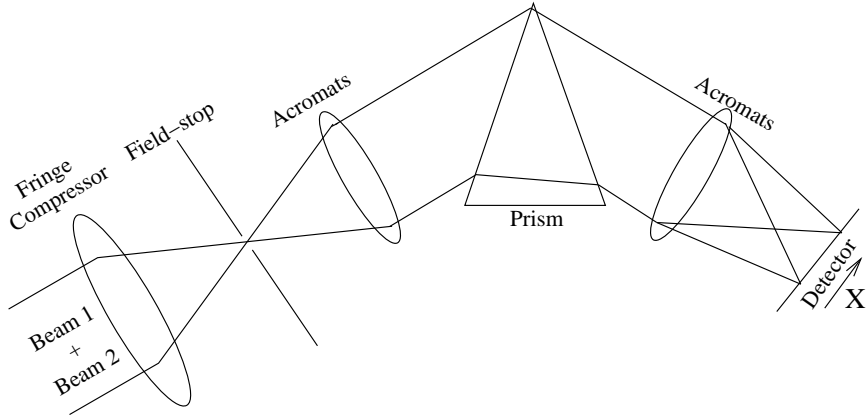}	
\caption{Schematic of the optics that follow the beam combiner and which are 
responsible for dispersing and imaging the combined light from the two 
collectors. The combined beams from the two siderostat collectors enter
at the left and proceed through the compressor and field stop to the prism 
disperser, and are finally imaged on the detector at right. \label{fig:optics}
}
\end{figure*}

We start by
considering how the light is processed after the beams from each arm of the
interferometer are combined. As described in Appendix A of Paper I (see
especially Figure 19), the collimated beams from each collector are passed
through the internal delay lines, and the two beams are superposed in the beam
combiner. The precise operation of these optical assemblies is not directly
relevant to our present goal, so we will not elaborate further on them here. The
combined beams then pass through the prism disperser, and are imaged on a
fast-readout charge-coupled device (CCD) in the focal plane of the camera.
Figure \ref{fig:optics} is a sketch of this optical system.
%
%

\begin{figure*}[!ht]
\epsscale{1.5}
\plotone{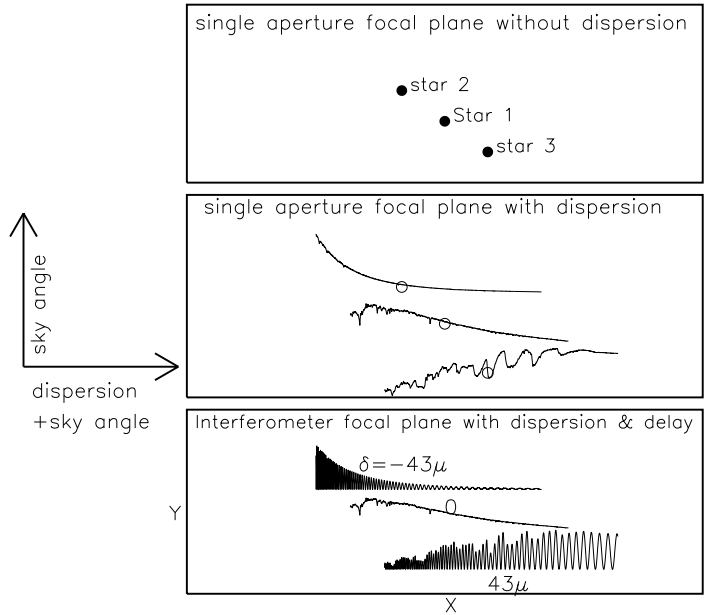}	
\caption {Schematic of the ``ideal" focal plane. \textbf{Top:} When the light
from one of the siderostats of SIM is blocked and the prism removed from the
light beam, the stars will appear as spots of light in this idealization. Stars
2 and 3 are offset from star 1 in this sketch by $\approx 1''$ in each
dimension. \textbf{Middle:} The star images will spread into spectra when the
prism is introduced into the path of the (single) beam. The three stars 1, 2,
and 3 shown for illustration in the field are of spectral type G, O, and M,
respectively; note how their SEDs differ. We have adopted the convention that
the short-wavelength end of the spectrum appears to the left (small $X$), and
the wavelength increases with position $X$ to the right; this is otherwise
arbitrary, and depends on the design of the instrument. For this sketch we have
further assumed that the relation between $X$ and $\lambda$ is simply linear.
\textbf{Bottom:} When the beam from the second siderostat is added, the spectra
will be modulated with a complicated fringe pattern, as described in the text.
If the white fringe of the instrument happens to coincide with one of the stars
(assumed to be star 1 here), the interferogram of that star will show no
fringes, but will reproduce the spectral energy distribution of the star. Stars
3 and 2 are at delay offsets of $\pm 43\, \mu$m from star 1, respectively.
\label{fig:focalplane}}
\end{figure*}

In order to understand the structure of the images which appear in SIM's focal 
plane, it is useful to proceed in a series of steps of increasing complexity. 
For the moment, we will assume that SIM's primary apertures are very large, so 
that star images appear as unresolved dots (this will be rectified later). 
Figure \ref{fig:focalplane} shows a sketch of the 2-D images which appear in 
such an ``ideal'' focal plane (at the location of the detector at the right
side of Figure \ref{fig:optics}).
\begin{itemize}
\item If one of the siderostats is covered and the dispersing prism removed from
the light beam, the image appearing at the focal plane will show the stars in
the FOV at their respective positions. In the top panel of Figure \ref
{fig:focalplane} we have sketched a field with 3 stars of different spectral
types in the FOV: \#\,1 is a G star at the center; \#\,2 an O star to the upper
left; and, \#\,3 an M star to the lower right.
\item Next, if we introduce the prism into the light path, the images of the
stars will be ``spread out'' in the direction of the dispersion (assumed to be
the $+X$ direction in Figure \ref{fig:optics}) in the same fashion as with an
\textit{objective prism spectrograph}; the focal plane will consist of
``patches'' of light, each varying according to the SED of its star, as shown
schematically in the middle panel of Figure \ref{fig:focalplane}\footnote{The
SEDs are shown here as graphs for simplicity; in fact they will appear as
elongated patches of light with intensity distributions in $+X$ as shown in the
graphs.}.
\item If we now uncover the second siderostat, each star's spectrum will become
an \textit{interferogram} consisting of its SED modulated by a complex pattern
of fringes depending on the setting of the internal delay and the location of
the star on the sky. This modulation pattern can be read off along vertical
lines in the \textit{fan diagram} of Figure \ref{fig:FANdiagram} (to be
described in the next section).
\item Finally, if we scan the internal delay $\delta$, each star's interferogram
will change in a different way, according to location of the star on the
$x$-axis of the fan diagram. If the white fringe happens to be positioned
exactly at the location of a star, the interferogram of that star will lose its
fringes and show simply the stellar SED.
\end{itemize}

As an aid to understanding the final \textit{point spread function} for SIM
which results from the steps described above, we recall the concept of the
\textit{fan diagram} which was introduced in Paper I.

\section{The Fan diagram}
\label{sec:fan_diagram}

\begin{figure*}[!ht]
\epsscale{1.5}
\plotone{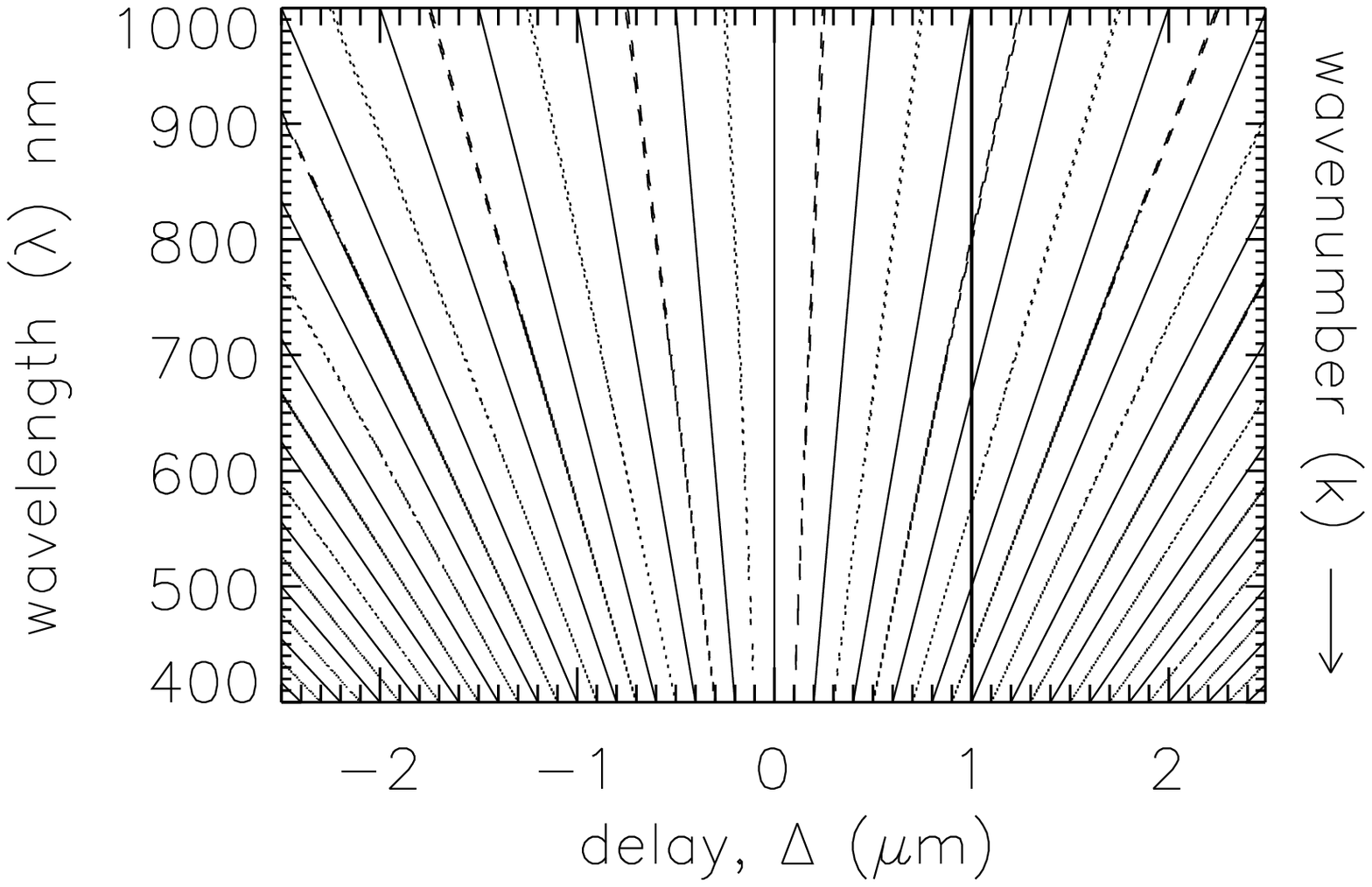} 
\caption {The ``fan diagram'', a contour plot of SIM's astrometric
interferometer response (the ``point spread function'', or PSF) as a function of
two variables: The optical path difference (OPD) $\Delta = \delta - B\theta$
between the internal and the external path delays, and; the wavelength
$\lambda$. The fringes appear to `fan out' at longer wavelengths, hence the
name. The point of (anti-)symmetry on the x-axis is called the \textit{white
fringe}. The vertical line at $\Delta =\delta - B\theta = +1 \, \mu$m traces the
interferometer's spectral response to a star located in the FOV at an angular
offset of $-0.023''$ from the white fringe. Scanning the internal delay $\delta$
will result in moving the location of this vertical line along the $x$ axis. An
``astrometric delay'' measurement with SIM begins by adjusting $\delta$ until
the star is situated at the white fringe. The dotted, solid and dashed contours
represent contour levels of 0, 1 and 2 respectively.
\label{fig:FANdiagram}
}
\end{figure*}

The response pattern $P(\delta, \theta, \overline{k})$ of SIM's astrometric  
(Michelson) interferometer in a quasi-monochromatic channel can be written as
(cf.\ Paper I):
\begin{equation} \label{eqn:SIMresponse}
P(\delta, \theta, \overline{k}) = P_{0} \{ 1 +
 A \sin ( 2 \pi \overline{k} [\delta - B \theta ] ) \},
\end{equation}
where $P_0$ is the total power, $A$ is the resultant fringe amplitude, 
$\overline{k}$ is mean wavenumber of the channel, $\delta$ is the
\textit{internal path delay}, and $B \theta$ is the \textit{external path delay} 
where $B$ is the projected baseline length and $\theta$ is an angle on the sky
measured from the direction perpendicular to the projection of the
interferometer baseline. A contour plot of this response pattern as a function
of the \textit{optical path difference} (OPD) $\Delta = \delta - B \theta$ and
$\lambda = 1/\overline{k}$ is shown in Figure~\ref {fig:FANdiagram}. It has the
appearance of a hand-held collapsable fan, and hence the name \textit{fan
diagram}\footnote{In their paper on observing binary stars with SIM,
\cite{dalalgriest01} introduce the ``channeled spectrum'', which is related to
the fan diagram defined here.}.
%
%
The fan diagram is a convenient and compact way of displaying the response of
SIM's interferometer in terms of delay and wavelength. Although it is not itself
a ``picture'' of SIM's focal plane, as we shall see it is an aid in constructing
such a picture for an arbitrary ensemble of target and field stars. A few
salient features of the fan diagram are:
\begin{enumerate}
\item  It is an intrinsic property of the instrument; it describes the \textit
{theoretical point spread function} of SIM, i.e.\ the idealized response to a 
point source at position $\theta$ at mean wave-number $\overline{k}$ and setting
$\delta$ of the internal delay.
\item A horizontal profile through the diagram shows a single fringe with a
period which increases gradually from short to long wavelengths.
\item The diagram shows an (anti-)symmetry about the zero point on the $x$ axis.
At this OPD, the fringe position is the same at all wavelengths, hence this
position is called the \textit{white fringe}. At larger $x$ offsets, the fringes
are increasingly tilted, and the ``blades of the fan'' become more and more
horizontal.
\item A star in the FOV is represented by a vertical line in the fan diagram. We
may think of this vertical line as shifting along the $x$ axis as the internal
delay $\delta$ is changed. The basic astrometric measurement procedure on a
single target star involves adjusting the internal delay $\delta$ until the
white fringe coincides with the target delay position, i.e.\ until the vertical
line is shifted to $\Delta = 0$.
\item The presence of multiple stars in the FOV will lead to a complicated
response as stars at different values of $\theta$ each contribute different
fringe patterns. The total response is the sum of all these patterns.
\end{enumerate}

\section{Modeling the real focal plane}

The real focal plane of SIM differs from the idealized model we have just
presented in a number of important ways: First, owing to diffraction by the
finite size of the collector, the images of stars will be large and will grow
(in size) with increasing wavelength. Second, the dispersion by the prism may
result in overlap of the diffracted images of different stars in a crowded
field, and these images will not overlap at the same wavelengths. Third, the
pixels of the CCD detector are much larger than those of our ideal model.
Finally, the outputs in each CCD pixel are binned as the internal delay is
scanned in order to measure the fringe parameters (total power, amplitude and
phase) in each wavelength channel. In this section we consider each of these
additional complications in turn.

To begin, we shall need a numerical model of the ideal focal plane which will
allow us to adequately represent the various smoothing and overlaps described
above. This model involves adopting a 2-D grid of data points with adequate
resolution in each direction. The various considerations which go into this
choice are discussed in Appendix \ref{app1}; we have settled on an array of size
$2112 \times 192$, and taken each element to represent the brightness in an area
of $1 \times 1 \, \mu$m in the focal  plane, corresponding to a square of side
1/12 arcsec on the sky for scale expected to be adopted in SIM's camera.

\subsection{Diffraction}

SIM's siderostat collectors will have effective outer and inner diameters of
304.5~mm and 178~mm, respectively. The point spread function will therefore be
approximately an Airy function with a full width at half maximum which increases
from $\approx$ 0.3\arcsec\ at 400~nm to $\approx$ 0.7\arcsec\ at 1000~nm. The
field of view (FOV) of the sky which appears on the detector is limited by the
field stop (Figure \ref{fig:optics}); the stop size is expected to be equivalent
to a circle of diameter $3''$ in the camera focal plane. Our model includes
light diffracted through this stop from all stars lying within a circle of
diameter 6\arcsec.

\subsection{Dispersion}
\label{subsec:dispersion}

\begin{figure*}[!ht]
\epsscale{2.4}
\plottwo{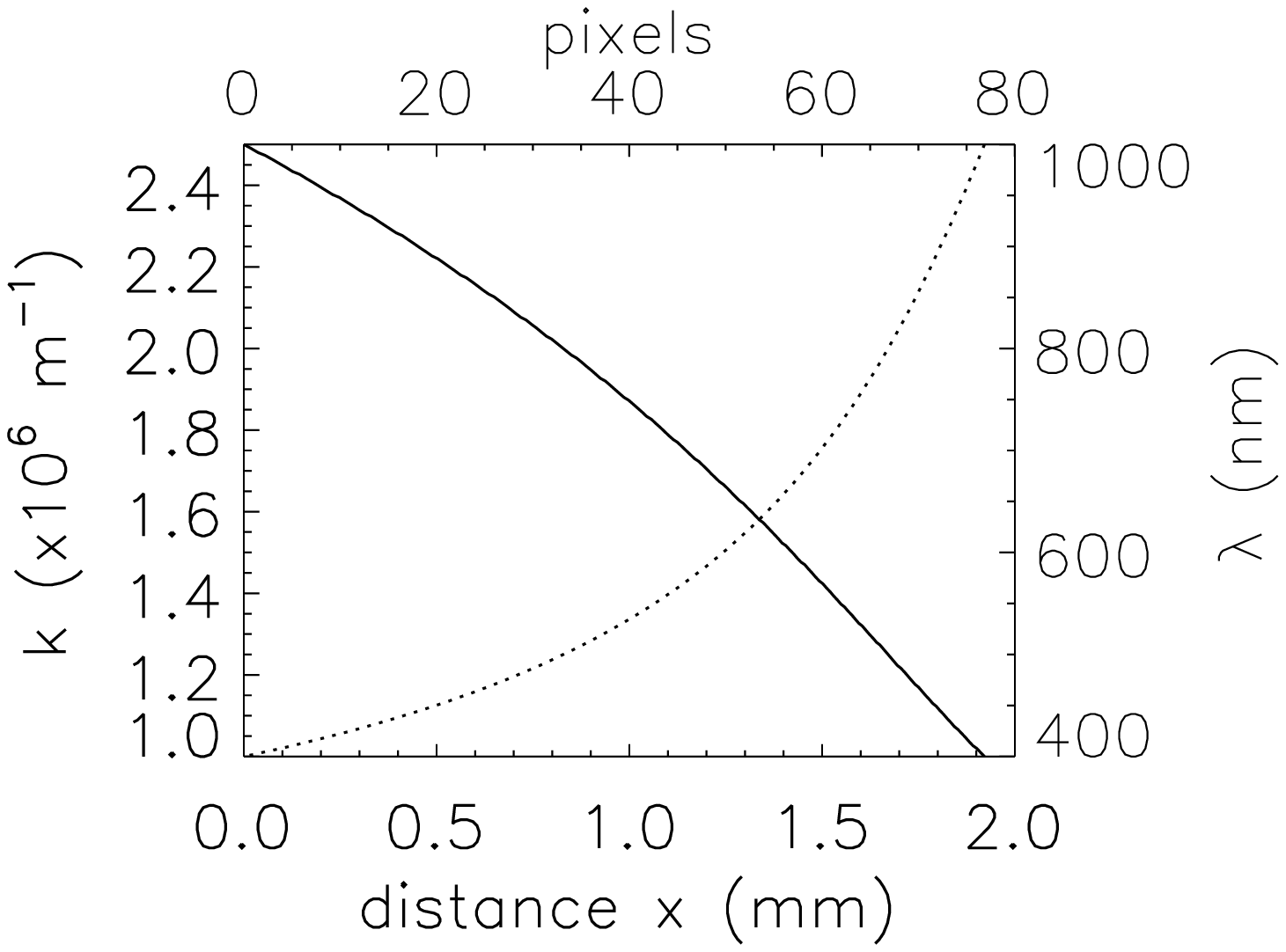}{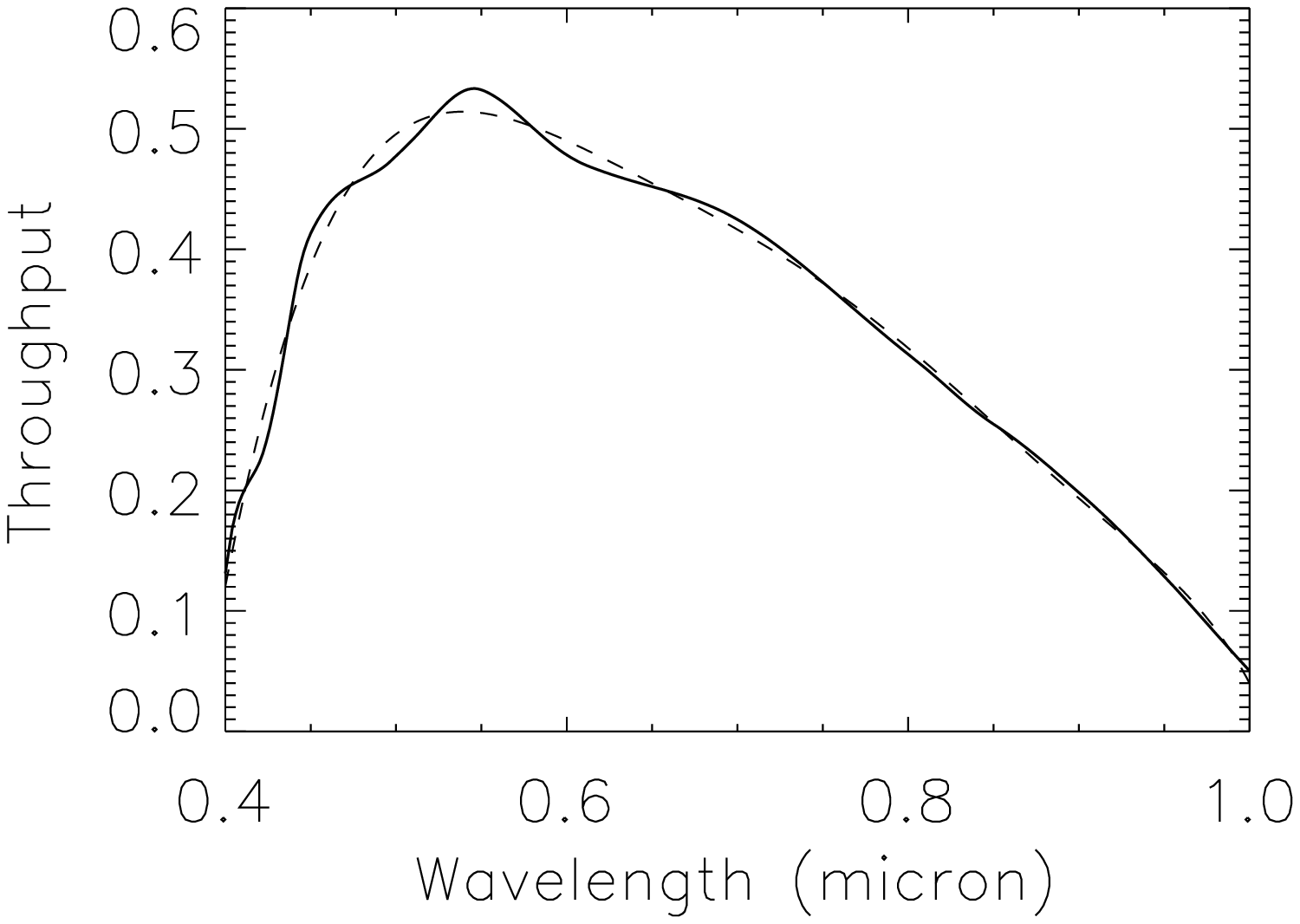}
\caption {\textbf{Left:} Experimentally-measured dispersion data for the
current design of SIM's prism disperser. The dotted line is for wavelength
(right axis) and the solid line for wavenumber (left axis). Note that the
dispersion is a non-linear function of wavelength $\lambda$, but close to a
linear function of wavenumber $k$.  \textbf{Right:} Estimated throughput
of the entire instrument including reflectivity of the optics and the detector
response. The dotted curve is a convenient polynomial fit to the measured values
(solid curve). \label{fig:DispThruput}
}
\end{figure*}

Next we must specify the prism dispersion, i.e.\ the relation between $\lambda$
and $X$. For a star at the center of the FOV, the prism used in SIM has a
measured dispersion as shown in the left panel of Figure \ref{fig:DispThruput}.
This relation can be closely approximated with a polynomial; the coefficients
are given in Appendix B of Paper I. It is likely that these coefficients will be
somewhat dependent on the location of the target star in the field of view;
however, these details are unknown at present, so this effect has not been
included in our models.
%
\begin{figure*}[!t]
\epsscale{2.2}
\plotone{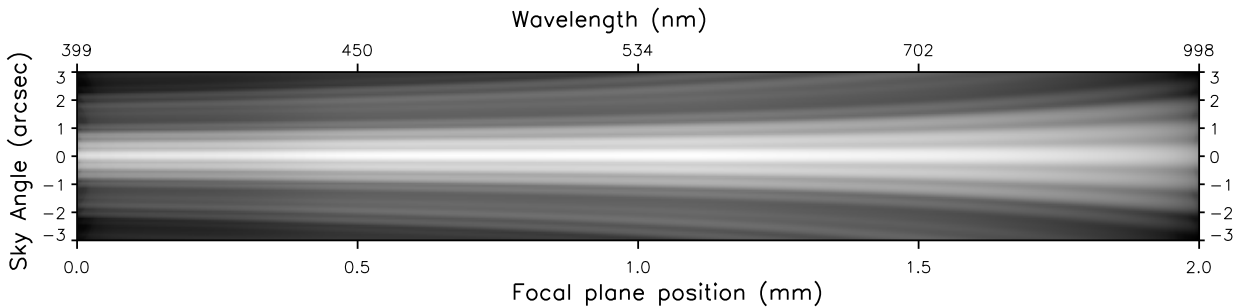}
\caption {Model of the ``ideal'' image in the focal plane of SIM for a single
target star, showing the combined effects of diffraction and dispersion. Note
that the size of the PSF increases as the wavelength increases in the $+X$
direction (horizontally to the right).  The intensity has been displayed on a
logarithmic scale in order to show the ``Airy rings'', which are no longer rings
at all, but long horizontal features which spread out with increasing
wavelength. For simplicity, a flat SED has been used in this illustration and
hence the intensity variation in the $+X$ direction mimics the throughput
variation. \label{fig:psf_variation}
}
\end{figure*}

The combined effects of diffraction and dispersion in SIM's focal plane are
shown in Figure \ref{fig:psf_variation}. Note that one result of this method of
dispersing the fringes is that each point in the focal plane receives photons
from a range of wavelengths of the target. The intensity at a given position in
the focal plane will change, as will the nominal wavelength associated with that
position, and the extent of this change will depend on the SED of the target.
This will also lead to small changes in the estimated fringe phase. Note further
that this effect is always present, even if there are no other stars within the
FOV of SIM; it is a ``self-confusion'' effect. We will quantify this effect for
SIM later in this paper.

\subsection{Throughput and fringe modulation}

There are two remaining modifications to the distribution of intensity in the
focal plane. The first is caused by the variation in the overall throughput of
the system as a function of wavelength, including the reflectivity of the optics
and the sensitivity of the CCD detector in the focal plane. The current best
estimate of this throughput is shown in the right panel of Figure
\ref{fig:DispThruput}. A polynomial fit to this function is given in Appendix B
of Paper I. As with the wavelength dispersion, the throughput is also expected
to be weakly field-dependent, but as yet we do not have sufficient knowledge to
model this effect.

The final (and in many ways most important) effect to include is the modulation
of the spectra by the fringe modulation function. This is \textit{strongly field
dependent}. This function will differ for each star in the FOV, and is given by
the fan diagram of Figure \ref{fig:FANdiagram}, which depends on the baseline
length $B$, the location $\theta$ of the star in the FOV, and the position of
the internal delay $\delta$.

A few clarifying remarks on the geometry are in order here. The relative angular
location of each star in the FOV is measured  as \textit{projected} onto the
interferometer baseline; see Figure 3 in Paper I for a sketch. The prism
disperser in SIM's optical train is oriented such that the direction of
dispersion is parallel to the projection of the baseline on the sky\footnote{It
could as well be perpendicular to the baseline, or at any other arbitrary
orientation.}, and the camera CCD is oriented such that the $+X$ pixel grid
direction is also aligned with the baseline projection. Both of these latter
choices are a matter of convenience and not of necessity. We have used them here
as well, since they simplify the model calculations.

\begin{figure*}[!t]
\epsscale{2.2}
\plotone{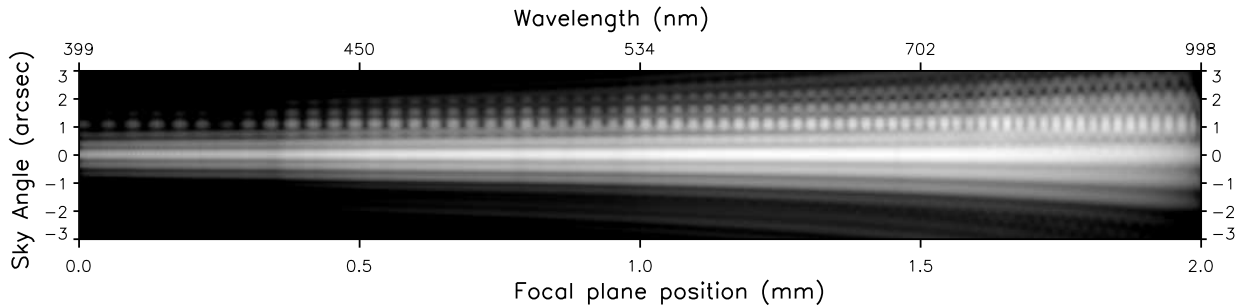}
\caption {A portion of the image in the ideal focal plane of SIM including all
effects discussed in this paper: diffraction, dispersion, throughput, fringe
modulation, and confusion. The FOV includes two stars, a G-type ``target'' star
at the center, and an M-type ``field  star'' offset by 1.5\arcsec\ towards the
upper left corner of the FOV. In this simulation, $\Delta=0$ on the target star,
so we see only the SED of this star (times the throughput, which falls off at
either end of the spectrum). The (faint) widely-spaced dark bands in this
spectrum are spectral features of this star. Above this G star is the spectrum
of the field M star which is shifted $\approx 1''$ vertically upwards and
$\approx 1''$ horizontally to the left of the target star and, since it is
offset substantially from the white fringe, it is also modulated by fringes
determined from the fan diagram. 
\label{fig:dispersed_spectra}}
\end{figure*}

Figure \ref{fig:dispersed_spectra} shows the ideal focal plane for two stars
separated by $\approx 1.5''$ and includes all effects discussed so far;
diffraction, dispersion, throughput, and fringe modulation. There is one aspect
of this figure (and also Figure \ref{fig:hrfocal_plane}) that may appear
paradoxical, and so deserves some clarification: We would expect the fringe
period (measured e.g.\ in microns) will be larger at the longer wavelengths,
which means that the fringes should become progressively more ``stretched out''
at larger $X$. However, it is clear that the fringes in Figures \ref
{fig:dispersed_spectra} and \ref{fig:hrfocal_plane} are actually more ``bunched
up'' at large $X$. The reason for this can be found in the dispersion curve of
Figure \ref{fig:DispThruput}; this curve shows that the same interval in $X$
covers a much larger range in wavelength at the red end of the spectrum than it
does at the blue end. The results are clear from the labelling of the $x$ axes
in Figures \ref{fig:psf_variation} and \ref{fig:dispersed_spectra}. For
instance, the $500\, \mu$ interval in $X$ from $0 \leq X \leq 0.5$ mm covers a
50 nm wavelength interval from $\approx 400 - 450$ nm at the blue end (left),
whereas the same interval from $1.5 \leq X \leq 2.0$ mm covers a wavelength
interval that is $\approx 6$ times larger at the red end (right), from $\approx
700 - 1000$ nm. It follows that in these figures the fringes will appear to
bunch together at large $X$, even though the intrinsic fringe period is a factor
of $\approx 2.5$ greater there.

\subsection{Delay scanning}

As described in Paper I, the fringe measurement process for SIM consists first
of mechanically adjusting a coarse component of the delay in order to position
the white fringe near to the target. The internal delay $\delta$ is then scanned
over a small range, typically $\pm 1 \lambda$, and data recorded at the various
wavelengths. This scanning of the delay leads to changes in the dispersed images
owing to the varying modulation of the fringe function.
%
\begin{figure*}[!ht]
\epsscale{1.5}
\plotone{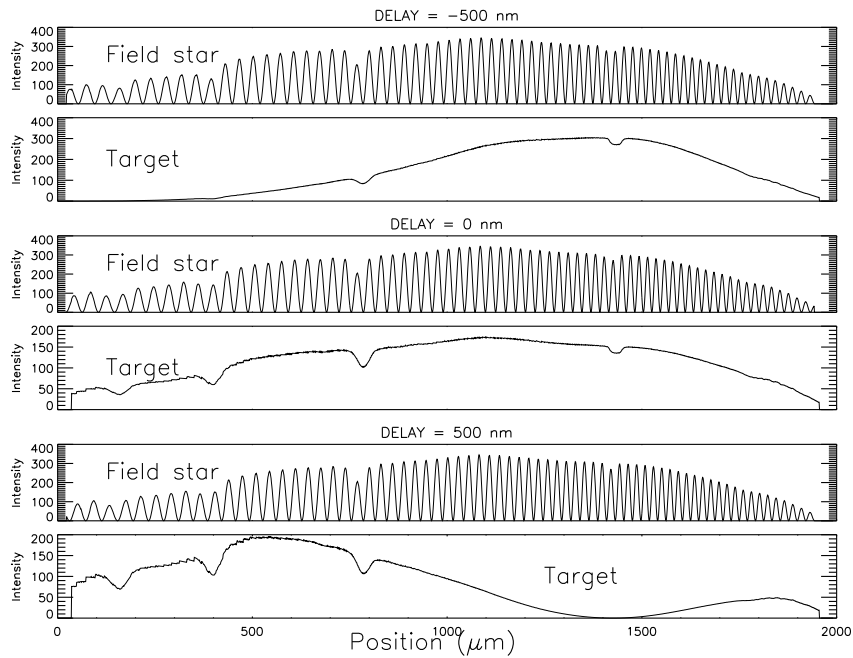}
\caption {Focal plane intensity distribution for three different values of the
internal delay. Only the central profiles in $X$ of the target and the field
star have been plotted. As the scanning progresses, both the target and the
field star show variations at differing rates, depending on their
(baseline-projected) locations in the FOV and the value of the internal delay.
The throughput and the dispersion have been included, but not the diffraction 
(which will reduce the fringe contrast, especially on the field star).
\label{fig:hrfocal_plane}}
\end{figure*}
%
Figure \ref{fig:hrfocal_plane} shows the dispersed images of the ``target'' and
the ``field star'' in Figure \ref{fig:dispersed_spectra} at 3 settings of the
internal delay, from -500 to +500 nm. For this simulation, the target star is
assumed to correspond to OPD = 0, so that the middle panel shows the SED of this
star. At the extremes of the internal delay, slow changes in the spectrum of the
target star are present, whereas the general appearance of the field star does
not seem to vary much. However, the resolution in the focal plane ($24 \, \mu$m,
see next section) is sufficiently fine that rapid variations can sometimes be
seen from one wavelength channel to the next.

\subsection{Binning}

The final step in our model of the ``real'' focal plane is to account for the
size of the CCD pixels, and for the integration in each CCD pixel during
scanning of the internal delay. Both these steps amount to \textit{binning}
during the acquisition of the data.

\subsubsection{Pixellation in the focal plane}

\begin{figure*}[!ht]
\epsscale{1.5}
\plotone{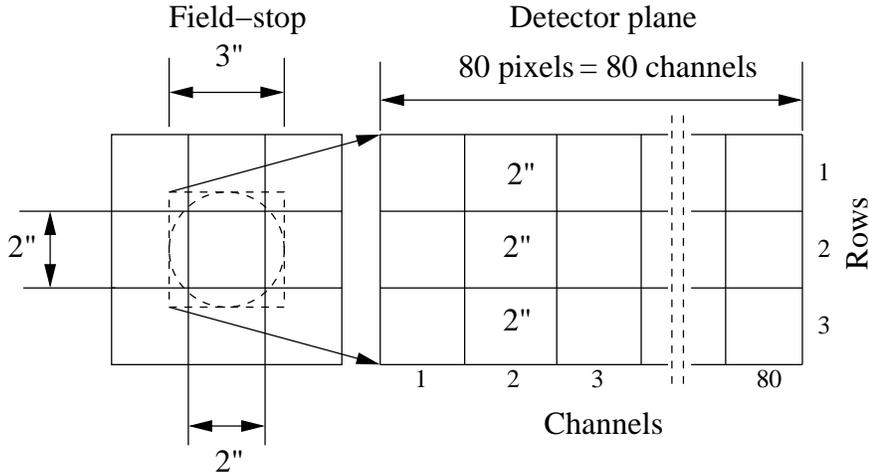}
\caption {Correspondence between the ideal focal plane, the field stop, and
the CCD pixels in the FOV. \textbf{Left:} The dashed circle shows the $3''$
FOV, covered with 2\arcsec \ CCD pixels.  \textbf{Right:} The diffracted image
is dispersed in the $+X$ direction over 80 pixels. \label{fig:detp}
}
\end{figure*}

The CCD in the focal plane will have $24 \times 24 \, \mu$m pixels, which we
model by binning the $1 \, \mu$m pixels of our idealized focal plane simulation.
The camera optics are such that the scale in the focal plane will be $2''$ per
CCD pixel of $24 \, \mu$m. Figure \ref{fig:detp} shows the mapping between the
intensity in the region of the $3''$ field stop and the CCD detector plane. The
idealized focal plane is binned to $3 \times 80$ pixels, which is the effective
sensitive area of the SIM CCD to be used for analysis of the fringes. The
current default is to carry out a summation of the data in rows 1, 2, and 3 for
each of the 80 channels, and to provide for a series of on-board binning
patterns of the data in the ``channel'' direction. A different approach of
reading out each row separately may confer an advantage in crowded fields; we
will have more to say about this later in this paper.
%

\subsubsection{Integration of the fringe signal}

In addition to the pixellation introduced in the previous section, the fringe
signal of our idealized focal plane is integrated over a finite delay range (and
time) in order to simulate the operation of the fringe detector. This
integration occurs during scanning of the internal delay. The scanning involves
moving a dither mirror back and forth with an amplitude of $1 \, \mu$m on top of
an additional fixed delay following a triangular-shaped wave with a nominal
frequency of 250 Hz\footnote{The integration time of the fringe detection system
can be lengthened simply by decreasing the scanning frequency. This facilitates
observations of faint targets with SIM.}, and simultaneously recording eight or
sixteen exposures of $250 \, \mu$s in each of all 80 channels. The total power
$P_0$, the resultant fringe amplitude $A$, and the resultant fringe phase $\phi$
are then determined for each channel using a least squares method. In our
simulation of this process we calculated idealized focal plane intensity
distributions for every nanometer over the internal delay range of $ -0.5 \leq
\delta \leq +0.5 \mu$m and binned them to obtain 8 samples, each corresponding
to a delay step size of exactly 125 nm. We then used the on-board algorithm
\citep{milbas02} in order to calculate the fringe parameters.

\section{A comparison of the models}

In Paper I, we applied our simplified model to a number of SIM Key Projects and
estimated the expected single-measurement astrometric bias arising from
confusion. In that paper, we also tabulated the differences with the results
using the improved model of the present paper for several representative cases.
Table \ref{table:comparisons} reproduces those results here for convenience.
We have already noted that the differences are not significant, so that
a good estimate of the presence and magnitude of the confusion bias can be
obtained without the additional complexity of the improved instrument
model. In this section we discuss the comparison between the two models
in more detail.

In Paper I we found that the magnitude of the confusion bias depends on several
factors: the relative brightnesses of the field stars with respect to the target
star (including any vignetting by the entrance aperture); the SEDs of the target
and field stars; and the separation of the field stars from the target star as
projected along the interferometer baseline. Since these parameters have not
changed with the advent of the improved instrument model, and since the
calculations have been carried out without the addition of any noise sources,
the results ought to be identical. However, closer inspection of Table
\ref{table:comparisons} shows small but systematic differences, as follows:
\begin{itemize}
\item The total confusion biases we give in Table \ref{table:comparisons}
for the sample fields computed using the improved model (column 4) are all
\textit{smaller} than those of the simple model (column 3), except for case 5
(where the difference is zero), and case 7 (where the improved model is worse by
2 \muas).
\item The largest (absolute) differences (column 5) occur for cases 2, 4,
and 1, in that order. Case 2 involves a blue Z=0 quasar and a very red field
star of type M. Case 4 has all the same instrument parameters as case 2, and
also involves a very red M field star, but the quasar is now at Z=2, i.e. it is
less blue. Case 1 is identical to case 2, but now the field star is a less-red
type A.
\end{itemize}

To summarize, the improved model generally predicts smaller astrometric biases
than does the simple model, and the largest differences between the models
appear to occur for cases involving the reddest field stars. We now look for
the explanation of these effects in the differences between the models.

\subsection{Differences in the models}

The improved model uses the same phasor description as the simple model, and
consideration of this description provides a basis for understanding the origin
of the differences. Figure 6 of Paper I shows an example of a confused field. In
that figure, the target phasor is shown as the thick black arrow pointing along
the +X (real) axis, and 10 additional faint confusing stars are scattered over
the field. The resulting phasor no longer points along the +X axis, but has
small errors both in amplitude and phase compared to the target. The phase error
causes the astrometric confusion bias. Now suppose we have just one confusing
field star; the only way to reduce the amount of the bias $\phi$ in that case is
to reduce the amplitude of the phasor representing that field star. So we must
look for any differences in the improved model which might result in reducing
the fringe amplitude of the field star.

The first, and perhaps most obvious, way of reducing the contribution of the
field star is to reduce the size of the pixels. The improved model effectively
has $2'' \times 6''$ square pixels, which is a factor 1.7 larger than the $3''$
round pixels of the simple model, so this unfortunately goes the wrong way.
However, with the exception of case 6, the angular separation between the target
and the field star is a small fraction ($\approx 50/300 = 0.17$) of the
collector PSF, so this effect is unlikely to contribute much. The exception is
case 6 in Table \ref{table:comparisons} (angular separation $1.5''$), but this has
the faintest field star (at 3 mag below the target) of all the cases in Table,
so its bias effect was negligible to start with\footnote{We will return later to
the suggestion that reducing the (effective)pixel size on the CCD detector can
be a way of reducing the confusion bias for some special cases.}.

The next difference between the models to consider is that part of the
``binning'' which involves integration of the fringe signal during the delay
scanning. However, this refinement is unlikely to have any effect at all, for
the following reason: The simple algorithm modeled the fringe as a
densely-sampled sinusoid, while the improved model uses a much coarser binning
of the fringe into eight segments. However, the algorithm used to compute the
fringe parameters is the same for both models; it is a standard approach which
explicitly accounts for the time binning using the fact that the signal is a
sinusoid of known period (e.g.\ \cite{milbas02}, \cite{catmil02}). This
refinement of the improved model should therefore have no effect.

Finally, there is the binning of the dispersed fringes which occurs owing to the
pixellation of the CCD in the focal plane. This remaining effect appears capable
of accounting for the differences we have observed. The situation can perhaps be
best understood by referring to Figure \ref{fig:hrfocal_plane}. This figure
shows that the period of the dispersed fringes is very long for the target star,
but much shorter for field stars. The example shown in this Figure is an extreme
case of course, with a projected separation of $\approx 1''$. After pixellation
by the CCD, the amplitude of the target star fringe which will be recorded in
any given pixel (channel) as the delay is scanned will hardly be affected,
because the period of the dispersed fringes is many dozens of pixels (of order
80 if the target is at the white fringe). But the periods of field stars are
shorter, and the extreme example shown in this figure has a period of about 1
pixel near the middle of the CCD. It's clear that the fringe amplitude measured
on this field star as the delay is scanned will be greatly reduced, indeed,
essentially zero in this example. Field stars not as distant from the target
will be less affected, but the amplitudes of their phasors will nevertheless be
smaller, leading to a reduction in the confusion bias they contribute. Note that
this effect will never lead to an increase in the amount of the confusion bias,
only a decrease, as is observed in Table \ref{table:comparisons}. To conclude
the argument, we note that the period of the dispersed fringes on the CCD (cf.
Figure \ref{fig:dispersed_spectra}) is larger at the blue end of the spectrum
(about 2 pixels in this example) than it is at the red end (about 0.8 pixels).
The smoothing effect of pixellation on the CCD will therefore be most severe for
the reddest field stars, as is also observed in Table \ref{table:comparisons}.
We conclude that it is the pixellation by the CCD in the focal plane which is
responsible for the reduction in the level of confusion bias recorded with the
improved instrument model described in this paper.

\begin{deluxetable}{ccccc}
\tablecaption{Comparison of the single-measurement astrometric bias obtained
with the improved instrument model to the results of the simplified model of
\citet{sriron07a}.\label{table:comparisons}}
\tablewidth{0pt}
\tablehead{
\colhead{Case} & \colhead{Model} & \colhead{Simple} & \colhead{Improved} &
\colhead{$\mid$Difference$\mid$} \\
\colhead{Number} & \colhead{parameters} & \colhead{(\muas)} & \colhead{(\muas)}&
\colhead{(\muas)}
 }
\startdata
1    & Q(z=0), A1V, 2, 50, 5, 90   & 442 &  431 &      11\\
2    & Q(z=0), M6V, 2, 50, 5, 90   &1072 &1036  &  36       \\
3    & Q(z=2), A1V, 2, 50, 5, 90  &224      & 215     &  9      \\
4    & Q(z=2), M6V, 2, 50, 5, 90  &614     & 592     & 22        \\
5    & Q(z=2), M6V, 2, 50, 5, 10 &-46    &  -46    &    0     \\
6    & A1V, B1V, 3, 1500, 10, 90  &-1     & 0    &     1   \\
7    & A1V, M6V, 2, 50, 90, 90  &-59     &-61      & 2       \\
8    & A1V, M6V, 2, 25, 90, 90  &-53     & -46     & 7       \\
\enddata
\tablecomments{In column 2, Q(z=2), A1V, 2, 50, 5, 90 refers to a model with
a redshift 2 target quasar and an A1V field star with $\Delta$m = 2, located
50 mas distant in PA = 5\degr. The baseline orientation is 90\degr.
Cases 1-5 are taken from the Quasar Frame Tie key project (cf. Paper I).
Cases 6-8 are taken from a selection of binary models.}
\end{deluxetable}

\section{Discussion}

A number of issues have been uncovered during the course of this work which
deserve further scrutiny, but for which there is insufficient space in this
paper to deal with properly. We briefly mention several of these remaining
issues.

\subsection{Separate row readout}

\begin{figure*}[!ht]
\epsscale{1.5}
\plotone{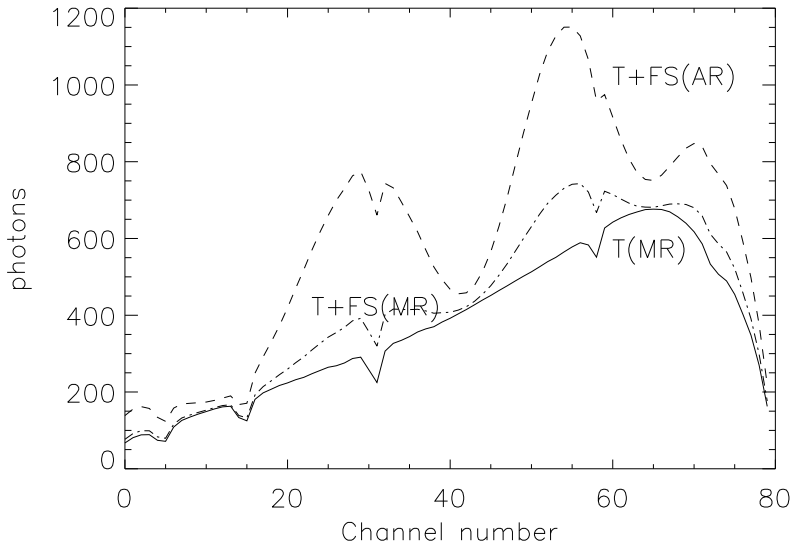}
\caption {T(MR) (solid line); signal spectrum when only
a single target star is present at the delay center in the FOV, and only the
middle row of pixels is used. T+FS(MR); as previously, but with the field star
present. T+FS(AR); as previously, but adding the signal in all 3 rows. Note how
the contribution from the field star has grown in the last plot, dominating the
signal around channels 30 and 55. \label{fig:row_wise_readout1}
}
\end{figure*}

\begin{figure*}[!ht]
\epsscale{1.5}
\plotone{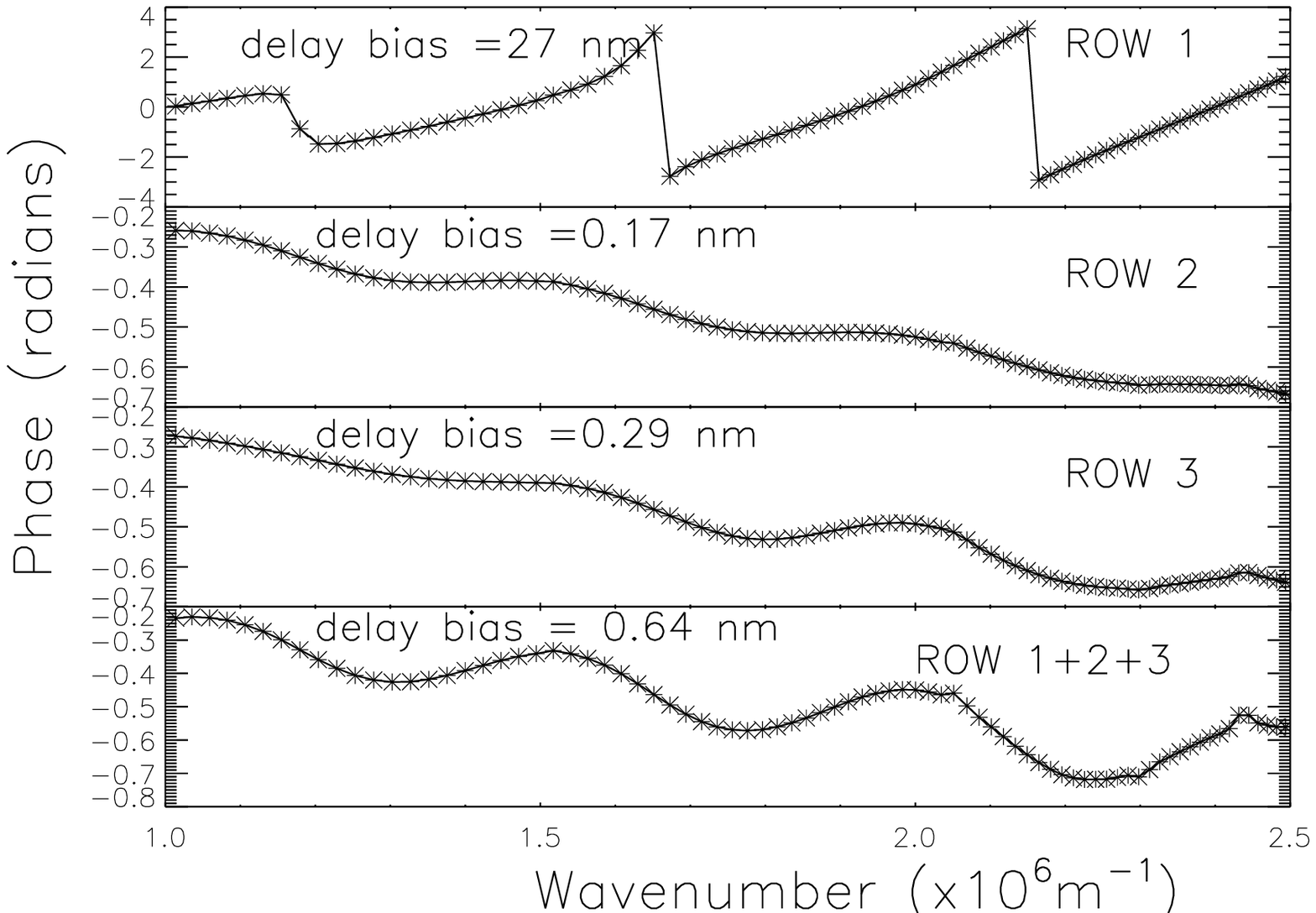}
\caption {Phase spectrum obtained separately from each of the three rows of
the CCD detector for the binary target described in the text, and finally
from the summed signal. In all 4 cases the bias in the astrometric delay
caused by the presence of the field star is included. The bias obtained
from adding the intensities of all the three rows (0.64~nm) is higher than
that determined using only the middle row(0.17~nm).
\label{fig:row_wise_readout2}
}
\end{figure*}

As explained earlier, (cf. Figure~\ref{fig:detp}), there will be three rows of
80 pixels on the detector devoted to collecting the dispersed fringe photons.
The nominal SIM design is to sum the charge from all 3 rows on board to form a
single row of 80 pixels; the fringe parameters will be estimated from these data
as the internal delay is scanned. Here we show that, in crowded fields, it may
be useful to read the rows out separately, and to estimate three sets of fringe
parameters, one from each row. For example, consider a field consisting of a
binary in which one of the stars (which we designate as the ``target'') is in
the middle pixel, and the other star (the ``field'' star) is in the upper pixel.
The field star in this simulation is fainter by 2 magnitudes, and located at a
radial distance of 1.4\arcsec\ at a position angle of 2\degr.  The projected
separation (along the interferometer baseline) is about 50 mas.
Figure~\ref{fig:row_wise_readout1} shows the signal spectrum on the detector
with and without the field star. This suggests that the astrometric bias arising
from confusion would be significantly reduced if only the central row of pixels
is used to compute the fringe parameters.
%
%
A comparison of the phase spectrum computed for each case confirms this
suggestion, as shown in Figure~\ref{fig:row_wise_readout2} (See Paper I for the
definitions and a discussion of the phase and delay spectra). The phase spectrum
computed using only the central row of pixels shows the least perturbation, and
the resulting bias in the astrometric delay caused by this field star is small
(0.17 nm). However, using all 3 rows results in a significant delay bias which
is almost 4 times larger. Clearly, in this case the best strategy would be to
use only the central row of pixels. In fact, looking for consistency in the
fringe parameters calculated from each row separately may be a useful indicator
of the presence of confusion, especially if the details of the distribution of
field stars is not known ahead of time. This subject clearly warrants a more
detailed discussion; it is possible that the advantages of reading out the data
for each row separately may outweigh any signal-to-noise penalty. In Paper I we
concluded that modeling and removing the astrometric bias arising from confusion
in SIM measurements is not likely to be feasible owing to the lack of
sufficiently-detailed SEDs and accurate positions of all the relevant stars in
the FOV. However, whatever level of bias may be present in the data, it is quite
possible that this bias can be further reduced by using the additional
flexibility offered by separately analyzing the 3 rows of the CCD detector.

\subsection{Self-confusion}

A description of how SIM can in principle `confuse itself' was given in \S
\ref{subsec:dispersion}. The basic problem here is that, even when there is only
one star within the FOV, the effective wavelengths of the channels will be
affected by photons diffracted into that channel from neighboring channels. This
effect will clearly depend on the spectrum of the target star, including e.g.\
absorption features. We find that there could be a maximum of 0.5\% change in
the effective wavelength of a given channel both for A and M type stars.
Accordingly, we estimated the fringe phase first with the actual wavelengths of
each channel, and then with a random perturbation of those wavelengths by 0.5\%.
The differences in the astrometric delay are 0.2~nm for an A type star and
0.35~nm for an M type star.  The corresponding single-measurement bias in the
position measurement is $\approx 8 \mu$as, which is likely to be negligible in
most cases.

The question arises as to just how the effective wavelength of each channel will
be calibrated in orbit. If a bright star is used for this purpose, the effects
of `self-confusion' described above will be included in the calibration.
However, it is also clear that the results of such a calibration will be
dependent on the SED of the calibrating star, so this choice will have to be
made carefully.

There is a variant on this problem when the FOV includes field stars as well as
the target star. In this case it is clear that the effective wavelength of the
sum of all the photons arriving at any position in the focal plane will depend
on the SEDs of all the field stars as well as that of the target star. Our
simulations indicate that this problem is slightly more severe than the
`self-confusion' described above, and in general the change in wavelength
arising from the presence of multiple sources is greater in the longer
wavelength channels. We have estimated this effect in a few test cases and find
the biases to be in the range of 1-10~nm (20-200~\muas) at short wavelengths,
and slightly more at long wavelengths. The pixellation of the focal plane tends
to smooth out this change in effective wavelength to some extent. Again, one way
at least to identify the presence of a possible problem is to estimate the
fringe parameters in each row separately; if they are inconsistent, it may help
to ignore the red end of the phase spectrum.

\cite{milcattur02} have modeled the path-delay error for SIM arising from
wavelength errors and shown it to be small. However, their analysis did not
include the diffraction effects we have described above.

\subsection{Loose ends}

Finally, we note that even our ``improved'' instrument model contains numerous
idealizations which deserve further scrutiny. For example, our model for the
pixellation of the focal plane assumes there is no space between the CCD
pixels, and the response across these pixels is uniform. Similarly, variations
of parameters across the FOV such as the throughput and the dispersion have
been ignored. 

\section{Conclusions}

We have developed an improved instrument model for SIM which includes a number
of important effects that were not considered in our first paper. The additional
features included here provide a much more detailed understanding of just how
SIM works. We note that this significantly-more-complicated model does not lead
to any major changes in our initial estimates of the astrometric bias arising
from extraneous field stars in the field of view (cf.\ Paper I), and the small
differences which have appeared for the sample fields we have calculated can be
explained as a consequence of smoothing by the pixellation of the images on the
CCD in the focal plane of the camera. Besides the obvious pedagogical value of
the improved model, it is likely to be a useful point of departure for future
investigations of even more subtle instrumental effects which may be present in
real SIM data. 

\acknowledgments

We are grateful to our colleagues in the SIM Science and Engineering Teams for
discussions about the potential effects of confusion on SIM astrometry. Special
thanks go to Mike Shao for sharing his detailed knowledge about the design of
SIM, and to him and Steve Unwin for their encouragement to include the effects
of the measurement process discussed here into our modeling of confusion as
described initially in Paper I. We have especially appreciated the constructive
criticisms of the referee on an earlier version of this paper. This work was
funded by the Jet Propulsion Laboratory under contract \#1268384, and carried
out at the Space Telescope Science Institute. We thank the Institute director
for providing additional financial support to permit the timely completion of
this project and publication of the results.

For queries on the IDL code developed for these studies please contact the
authors.

\appendix
\section{Modeling the SIM focal plane}
\label{app1}

We have chosen to model the ideal focal plane with a $X,Y$ rectangular grid of
$2112 \times 192$ elements, based on the following considerations:

\begin{enumerate}
\item The spatial sampling at the focal plane was chosen for convenience to be
1$\, \mu$m. The CCD camera will have $24 \times 24 \, \mu$m square pixels, each
of angular size $2'' \times 2''$ on the sky.  The linear size of each array
element in our model of the focal plane therefore corresponds to an angular size
of $1/12''$.
\item In order to include the diffracted light from a star located as much as
$3''$ from the center of the FOV, its point spread function was simulated with a
separate array of grid size $120 \times 120$ elements, corresponding to a total
angular size of $10'' \times 10''$.
\item We have required that the light from all the stars within a circle of
diameter 6\arcsec\ centered on the FOV be included in the simulation. This
implies that the minimum size of the simulated FOV should be $6'' \times 6''$. 
Accounting for the size of the point spread function, the required simulation
area grows to $16'' \times 16''$. The simulated focal plane must therefore be
$16 \times 12 = 192$ elements in size along the direction normal to that of the
dispersion (the Y-direction).
\item The design of the camera was such as to provide 80 wavelength channels,
i.e.\ 80 CCD pixels along the direction of the dispersion. This corresponds to a
linear size of 1920 microns.
\item The spectra of stars not at the center of the FOV will be shifted along
the direction of dispersion. This may require as much as an additional $\pm 36$
elements in order to include all the relevant stars within the circle of
diameter 6\arcsec. Thus, the required number of array elements along the
direction of dispersion in $+X$ is $1920 + 2 \times 36 + 120 = 2112$.
\end{enumerate}


\end{document}